\documentclass[prd,twocolumn,a4paper,aps,showpacs,
               nofootinbib,nobibnotes,superscriptaddress]
              {revtex4}
\usepackage{epsfig}

\newcommand{\ETAL}{{\it et al.}}
\newcommand{\EG}{e.g.}
\newcommand{\IE}{i.e.}
\newcommand{\CF}{cf}

\newcommand{\CRIT}{{\rm crit}}
\newcommand{\TOT}{{\rm tot}}
\newcommand{\MAT}{{\rm m}}
\newcommand{\CDM}{{\rm c}}
\newcommand{\BAR}{{\rm b}}

\newcommand{\SCAL}{{\rm S}}

\newcommand{\EFF}{{\rm eff}}
\newcommand{\GAL}{{\rm g}}

\newcommand{\UUNIT}[2]{
{\;\mathrm{#1}^{#2}} }

\newcommand{\AD}{{\rm AD}}

\newcommand{\CI}{{\rm CI}}
\newcommand{\NID}{{\rm NID}}
\newcommand{\NIV}{{\rm NIV}}

\newcommand{\ep}{\epsilon}
\newcommand{\Ga}{\Gamma}
\newcommand{\OLa}{\Omega_{\Lambda}}
\newcommand{\La}{{\Lambda}}
\newcommand{\Om}{\Omega}
\newcommand{\om}{\omega}
\newcommand{\si}{\sigma}
\newcommand{\OO}{{\cal O}}
\newcommand{\FG}[1]{Fig.~#1}
\newcommand{\pars}{\mathbf{\Theta}}
\newcommand{\Li}{\mathcal{L}}
\newcommand{\PD}{\mathcal{P}_n}
\newcommand{\On}{\mathcal{SO}_n}

\newcommand{\be}{\begin{equation}}
\newcommand{\ee}{\end{equation}}
\def\gsim{\raise 0.4ex\hbox{$>$}\kern -0.7em\lower 0.62
ex\hbox{$\sim$}}
\def\lsim{\raise 0.4ex\hbox{$<$}\kern -0.8em\lower 0.62
ex\hbox{$\sim$}}
\newcommand{\lims}[2]{_{#1}^{#2}}

\begin{document}

\title{The cosmological constant and general isocurvature initial
conditions}

\author{R.~Trotta}
\email{roberto.trotta@physics.unige.ch}
\affiliation{D\'epartement de Physique Th\'eorique, Universit\'e de
  Gen\`eve, 24 quai Ernest Ansermet, CH--1211 Gen\`eve 4, Switzerland}

\author{A.~Riazuelo}
\email{riazuelo@spht.saclay.cea.fr}
\affiliation{Service de Physique Th\'eorique,
             CEA/DSM/SPhT, Unit{\'e} de recherche associ\'ee au
             CNRS, CEA/Saclay F--91191 Gif-sur-Yvette c\'edex,
             France}

\author{R.~Durrer}
\email{ruth.durrer@physics.unige.ch}
\affiliation{D\'epartement de Physique Th\'eorique, Universit\'e de
  Gen\`eve, 24 quai Ernest Ansermet, CH--1211 Gen\`eve 4, Switzerland}

\date{27 November 2002}

\begin{abstract}

We investigate in detail the question whether a non-vanishing
cosmological constant is required by present-day cosmic microwave
background and large scale structure data when general
isocurvature initial conditions are allowed for. We also discuss
differences between the usual Bayesian and the frequentist
approaches in data analysis. We show that the COBE-normalized
matter power spectrum is dominated by the adiabatic mode and
therefore breaks the degeneracy between initial conditions which
is present in the cosmic microwave background anisotropies. We
find that in a flat universe the Bayesian analysis requires $\OLa
\neq 0$ to more than $3 \si$, while in the frequentist approach
$\OLa = 0$ is still within $3 \si$ for a value of $h \le 0.48$.
Both conclusions hold regardless of initial conditions.
\end{abstract}

\maketitle

\section{Introduction}
Ever since the beginning of modern cosmology, one of the most
enigmatic ingredients has been the cosmological constant. Einstein
introduced it to find static cosmological solutions (which are,
however, unstable)~\cite{Einst}. Later, when the expansion of the
Universe had been established, he called it his ``greatest
blunder''.

In relativistic quantum field theory, for symmetry reasons the
vacuum energy momentum tensor is of the form $\ep g_{\mu \nu}$ for
some constant energy density $\ep$. The quantity $\La = 8 \pi G
\ep$ can be interpreted as a cosmological constant. Typical values
of $\ep$ expected from particle physics come, \EG, from the
super-symmetry breaking scale which is expected to be of the order
of $\ep \gsim 1 \UUNIT{TeV}{4}$ leading to $\La \gsim 1.7 \times
10^{-26} \UUNIT{GeV}{2}$, and corresponding to $\OLa \gsim
10^{58}$. Here we have introduced the density parameter $\OLa
\equiv \ep / \rho_\CRIT = \La / (8 \pi G \rho_\CRIT)$, where
$\rho_\CRIT = 8.1 \times 10^{-47} \,h^2 \UUNIT{GeV}{4}$ is the
critical density and the fudge factor $h$ is defined by $H_0 = 100
\,h \UUNIT{km}{} \UUNIT{s}{-1} \UUNIT{Mpc}{-1}$, it lies in the
interval $0.5 \lesssim h \lesssim 0.8$. $H_0$ is the Hubble
parameter today.

Such a result is clearly in contradiction with kinematical
observations of the expansion of the universe, which tell us that
the value of $\Om_\TOT$, the density parameter for the total
matter-energy content of the universe, is of the order of unity, $
\OO({\Om_\TOT}) \sim 1$. For a long time, this apparent
contradiction has been accepted by most cosmologists and particle
physicists, with the hind thought that there must be some deep,
not yet understood reason that vacuum energy --- which is not felt
by gauge-interactions --- does not affect the gravitational field
either, and hence we measure effectively $\La = 0$.

This slightly unsatisfactory situation became really disturbing a
couple of years ago, as two groups, which had measured luminosity
distances to type Ia supernovae, independently announced that the
expansion of the universe is accelerated in the way expected in a
universe dominated by a cosmological constant~\cite{Riess,Perl}.
The obtained values are of the order $\OO(\Om_\MAT) \sim \OO(\OLa)
\sim 1$ and cannot be explained by any sensible high energy
physics model. Tracking scalar fields or
quintessence~\cite{quint,quint2} and other similar
ideas~\cite{kess} have been introduced in order to mitigate the
smallness problem --- \IE, the fact that $\ep \sim 10^{- 46}
\UUNIT{GeV}{4}$. However, none of those is completely successful
and really convincing at the moment.

After the supernovae~Ia results, cosmologists have found many
other data-sets which also require a non-vanishing cosmological
constant. The most prominent fact is that anisotropies in the
cosmic microwave background (CMB) indicate a flat universe,
$\Om_\TOT = \Om_\MAT +\OLa = 1$, while measurements of clustering
of matter, \EG, the galaxy power spectrum, require $\Ga \equiv h
\Om_\MAT \simeq 0.2$. But also CMB data alone, with
some reasonable lower limit on the Hubble parameter, like $h >
0.6$, have been reported to require $\OLa > 0$ at high
significance (see, \EG, Refs.~\cite{Net,VSA} and others).

This cosmological constant problem is probably the greatest enigma in
present cosmology. The supernova results are therefore under detailed
scrutiny. For example Ref.~\cite{Mez} is not convinced that the data
can only be understood by a non-vanishing cosmological
constant. Cosmological observations are usually very sensitive to
systematic errors which are often very difficult to
discover. Therefore, in cosmology an observational result is usually
accepted by the scientific community only if several independent
data-sets lead to the same conclusion. But this seems exactly to be
the case for the cosmological constant.

This situation prompted us to investigate in detail whether
present structure formation data does require a cosmological
constant. One may ask whether enlarging the space of models for
structure formation does mitigate the cosmological constant
problem. There are several ways to enlarge the model space,
\EG~one may allow for features in the primordial power spectrum,
like a kink~\cite{subir}. In the present paper we study the
cosmological constant problem in relation to the initial
conditions for the cosmological perturbations. In a first step we
re-discuss the usual results obtained assuming purely adiabatic
models and we investigate to which extent CMB alone or CMB and
large scale structure (LSS) require $\OLa \neq 0$ in a flat
universe. We shall first present the usual Bayesian analysis, but
we also discuss the results which are obtained in a frequentist
approach. We find that even if $\Om_\La = 0$ is excluded at high
significance in a Bayesian approach this is no longer the case
from the frequentist point of view. In other words the probability
that a model with vanishing $\OLa$ leads to the present-day
observed CMB and LSS data is not exceedingly small. We then study
how the results are modified if we allow for general isocurvature
contributions to the initial conditions~\cite{BMT1,BMT2,TRD1}. In
this first study of the matter power spectrum from general
isocurvature modes we discover that a COBE-normalized matter power
spectrum reproduces the observed amplitude only if it is highly
dominated by the adiabatic component. Hence the isocurvature modes
cannot contribute significantly to the matter power spectrum and
do not lead to a degeneracy in the initial conditions for the
matter power spectrum when combined with CMB data. This is the
main result of our paper.

The paper is organized as follows: In the next section we discuss
the setup for our analysis, the space of cosmic parameters and of
initial conditions, and we recall the difference between Bayesian
and frequentist approach. In Sec.~3 we present the results for
adiabatic and for mixed (adiabatic and isocurvature)
perturbations. Sec.~4 is dedicated to the conclusions.

\section{Analysis setup}

\subsection{Cosmological parameters}

As it has been discussed in the literature, the recent data-sets,
BOOMERanG~\cite{Net}, MAXIMA~\cite{MAX2}, DASI~\cite{DASI},
VSA~\cite{VSAdat}, CBI~\cite{CBI} and Archeops~\cite{Arch} are in
very good agreement up to the third peak in the angular power
spectrum of CMB anisotropies, $\ell \sim 1000$. In our analysis we
therefore use the COBE data~\cite{COBE} (7 points excluding the
quadrupole) for the $\ell$ region $3 \leq \ell \leq 20$ and the
BOOMERanG data to cover the higher $\ell$ part of the spectrum (19
points in the range $100 \leq \ell \leq 1000$). Since Archeops has
the smallest error bars in the region of the first acoustic peak,
we also include this data-set (16 points in the range $15 \leq
\ell \leq 350$). Including any of the other mentioned data does
not influence our results significantly. The BOOMERanG and
Archeops absolute calibration errors ($10\%$ and $7\%$ at $1\si$,
respectively) as well as the uncertainty of the BOOMERanG beam
size are included as additional Gaussian parameters, and are
maximized over.  We make use of the Archeops window functions
found in Ref.~\cite{ARCH:homepage}, while for BOOMERanG a top-hat
window is assumed.

For the matter power spectrum, we use the galaxy-galaxy power
spectrum from the 2dF data which is obtained from the redshift of
about $10^5$ galaxies~\cite{2dF}. We include only the 22
decorrelated points in the linear regime, \IE, in the range $0.017
\leq k \leq 0.314 \quad [h \UUNIT{Mpc}{-1}]$, and the window
functions of Ref.~\cite{2dF} which can be found on the homepage of
M. Tegmark~\cite{teg}.

Our grid of models is restricted to flat universes and we assume
purely scalar perturbations. Since the goal of this paper is more
to make a conceptual point than to consider the most generic
model, we fix the baryon density to the BBN preferred value
$\Om_\BAR h^2 \equiv \om_\BAR = 0.020$~\cite{burles}. We
investigate the following 3 dimensional grid in the space of
cosmological parameters: $0.80 \leq n_\SCAL~(0.05) \leq 1.20$,
$0.35 \leq h~(0.025) \leq 1.00$, $0.00 \leq \OLa~(0.05) \leq
0.95$, where $n_\SCAL$ is the scalar spectral index and the
numbers in parenthesis give the step size we use. The total matter
content $\Om_\MAT \equiv \Om_\CDM + \Om_\BAR$ is $\Om_\MAT = 1 -
\OLa$, and $\Om_\CDM$ indicates the cold dark matter contribution.
For all models the optical depth of reionization is $\tau = 0$ and
we have 3 families of massless neutrinos. For each point in the
grid we compute the ten CMB and matter power spectra, one for each
independent set of initial conditions (see Sec.~\ref{Icsec} below).

\subsection{Allowing for isocurvature modes} \label{Icsec}

We enlarge the space of models by including all possible
isocurvature modes. As it has been argued in Ref.~\cite{BMT1},
generic initial conditions for a fluid consisting of photons,
neutrinos, baryons and dark matter allow for five relevant modes,
\IE, modes which remain regular when going backwards in time.
These are the usual adiabatic mode (AD), the cold dark matter
isocurvature mode (CI), the baryon isocurvature mode (BI), the
neutrino isocurvature density (NID) and neutrino isocurvature
velocity (NIV) modes. The CMB and matter power spectra from the
cold dark matter and the baryon isocurvature modes are identical
(see the argument given in Ref.~\cite{lewis2}) and therefore from
now on we will just consider one of those, namely the CI mode. We
assume that all these four modes are present with arbitrary
initial amplitude and arbitrary correlation or anti-correlation.
The only requirement is that their superposition must be a
positive quantity, since the $C_\ell$'s and matter power spectrum
are quadratic and thus positive observables. For simplicity we
restrict ourselves to the case where all modes have the same
spectral index. Initial conditions are then described by the
spectral index $n_\SCAL$ and a positive semi-definite $4 \times 4$
matrix, which amounts to eleven parameters instead of two in the
case of pure AD initial conditions. More details can be found in
Ref.~\cite{TRD1}. For the search among initial conditions we use
the simplex method described in Ref.~\cite{TRD1}, with the
following modification. We find it convenient to express the
matrix $\mathbf{A}$ describing the initial conditions as
\begin{equation}
\mathbf{A} = \mathbf{U} \mathbf{D} \mathbf{U} ^{T} ,
\label{e:A_mat}
\end{equation}
where $\mathbf{A} \in \PD$, $\mathbf{U} \in \On$, $\mathbf{D} =
\textrm{diag}(d_1, d_2, \dots, d_n)$ and $d_i \geq 0$, $i \in
\{1,2,\dots,n\}$. Here $\PD$ denotes the space of $n \times n$
real, positive semi-definite, symmetric matrices and $\On$ is the
space of $n \times n$ real, orthogonal matrices with $\det = 1$.
As explained above, here we take $n = 4$. We can write
$\mathbf{U}$ as an exponentiated linear combination of generators
$\mathbf{H}_i$ of $\On$:
\begin{equation}
\mathbf{U} = \exp\left(\sum_{i = 1}^{(n^2-n)/2}
                         \alpha_i{\mathbf{H}_i} \right) ,
\label{e:U1}
\end{equation}
with
\begin{equation}
\mathbf{H}_1
 = \left( \begin{array}{cccc}
          0 & 1 & 0 & \ldots \\
          -1 & 0 & 0 & \ldots \\
          0 & 0 & 0 & \ldots \\
          \vdots & \vdots &  \vdots & \ddots
          \end{array} \right) ,
\label{e:H1}
\end{equation}
and so on, with $- \pi / 2 < \alpha_i < \pi / 2$, $i \in
\{1,2,\dots, (n^2 - n) / 2 \}$. In analogy to the Euler angles in
three dimensions, we can re-parameterize $U$ in the form
\begin{equation}
\mathbf{U}
 = \prod_{i = 1}^{(n^2-n)/2} \exp \left( \psi_i{\mathbf{H}_i} \right) ,
\label{e:U2}
\end{equation}
with some other coefficients $- \pi / 2 < \psi_i < \pi / 2$, $i
\in \{1,2,\dots, (n^2 - n) / 2 \}$, whose functional relation with
the $\alpha_i$'s does not matter.  The diagonal matrix
$\mathbf{D}$ can be written as
\begin{equation}
\mathbf{D}
 = \textrm{diag}\left( \tan(\theta_1), \ldots, \tan(\theta_n) \right) ,
\label{e:D1}
\end{equation}
with $0 \leq \theta_i < \pi / 2$, for $i \in \{1,2,\dots,n\}$.  In
this way, the space of initial conditions for $n$ modes is
efficiently parameterized by the $(n^2 + n) / 2$ angles $\theta_i,
\psi_j$. In our case, $n = 4$ and the initial conditions are
described by the ten dimensional hypercube in the variables
$(\theta_1, \ldots, \theta_4, \psi_1, \ldots, \psi_6)$. This is of
particular importance for the numerical search in the parameter
space. One can then go back to the explicit form of $\mathbf{A}$
using Eqs.~(\ref{e:U2}),~(\ref{e:D1}) and (\ref{e:A_mat}).

For a given initial condition determined by a positive
semi-definite matrix ${\bf A}$ and a spectral index $n_\SCAL$ we
quantify the isocurvature contribution to the CMB anisotropies by
the parameter $\beta$ defined as
\begin{equation}
\beta \equiv \frac{\displaystyle
                   \sum_{\scriptscriptstyle X = \CI, \NIV, \NID}
                   \left<(\ell(\ell+1))C_{\ell}^{\scriptscriptstyle (X,X)}
                   \right>_\ell}
                  {\displaystyle \sum_{\scriptscriptstyle
                   Y = \AD, \CI, \NIV, \NID}
                   \left< \ell (\ell + 1) C_\ell^{\scriptscriptstyle (Y,Y)}
                   \right>_\ell} ,
\label{e:beta}
\end{equation}
where the average $<,>$ is taken in the $\ell$ range of interest,
in our case $3 \leq \ell \leq 1000$, and where
$C_{\ell}^{\scriptscriptstyle (X,X)}$ stands for the
auto-correlator of the CMB anisotropies with initial conditions
$X$.

\subsection{Bayesian or frequentist?}

For the sake of clarity, we briefly recall two possible points of
view which one can take when doing data analysis, the Bayesian and
the frequentist approach, and highlight their difference. More
details can be found, \EG, in Refs.~\cite{FC, redbook, KS}.
Another possibility is based on Markov Chain Monte Carlo
techniques, which we do not discuss here; see instead \cite{lewis}
and references therein.

When fitting experimental data, we minimize a $\chi^2$ over the
parameters which we are not interested in.  This procedure is
equivalent to marginalization if the random variables are Gaussian
distributed. The Maximum Likelihood (ML) principle states that the
best estimate for the unknown parameters $\pars$ is the one which
maximizes the likelihood function:
\begin{equation}
\Li = \Li_0 \exp(-\chi^2/2).
\label{e:Li}
\end{equation}
We then draw $1 \si$, $2 \si$ and $3 \si$ {\em likelihood
contours} around the ML point, \IE, the one for which the $\chi^2$
is minimal in our grid of models. The likelihood contours are
defined to be $\Delta \equiv \chi^2 - \chi^2_{\rm ML} = 2.30$,
$6.18$, $11.83$ away from the ML value for the joint likelihood in
two parameters, $\Delta = 1$, $4$, $9$ for the likelihood in only
one parameter. This is the Bayesian approach: in a somewhat fuzzy
way, likelihood intervals measure our degree of belief that the
particular set of observations used in the analysis is generated
by a parameter set belonging to the specified interval. In this
case, one implicitly accepts the ML point in parameter space as
the true value, while points which are further away from it are
less and less ``likely'' to have generated the measurements. This
is the content of Bayes' Theorem, which allows us to interpret the
joint conditional probability $\Li (\mathbf{x} \vert \pars)$ of
measuring $\mathbf{x}$ for a fixed set of parameters $\pars$ as
the inverse probability $P (\pars \vert \mathbf{x})$ for the value
of $\pars$ given the measurements $\mathbf{x}$.

On the other hand, in the frequentist approach one asks a
different question: What is the probability of obtaining the
experimental data at hand, if the Universe has some given
cosmological parameters, \EG, a vanishing cosmological constant?
Clearly, if we want to answer the question whether a certain set
of experimental data forces a non-vanishing cosmological constant,
this is actually the correct question to ask. To the extent to
which the $C_\ell$'s can be approximated as Gaussian variables,
the quantity $\chi^2$ is distributed according to a chi-square
probability distribution with $F = N - M$ degrees of freedom
(dof), which we denote by $P^{(F)} (\chi^2)$, where $N$ is the
number of {\em independent} (uncorrelated) experimental data
points and $M$ is the number of fitted parameters. From the
distribution $P^{(F)}$ one can readily estimate {\em confidence
intervals}. For a given parameter set $\tilde{\pars}$ with
chi-square $\tilde{\chi}^2$ the probability that the observed
chi-square will be larger than the actual value by chance
fluctuations is
\begin{equation}
\int_{\tilde{\chi}^2}^{\infty} P^{(F)}(x) dx \equiv 1 - \gamma.
\label{e:Pfp2}
\end{equation}
In other words, if the measurements could be repeated many times
in different realizations of our universe, the estimated
confidence interval would asymptotically include the true value of
the parameters $100 \gamma \%$ of the time.

It is customary in cosmological parameter estimation to present
likelihood plots drawn using the Bayesian approach. It is
misleading that such Bayesian contours are usually called
``confidence contours'', which properly designate {\em
frequentist} contours. Likelihood (Bayesian) contours are usually
much tighter than the confidence contours drawn from the
frequentist point of view. This is a consequence of the ML point
having often a $\chi^2 / F$ much smaller than $1$, because the
data-sets are highly consistent with each other and also because
usually not all points are completely independent. If we consider
the usual situation in which likelihood contours are drawn in a
two dimensional plane with all other parameters maximized, the
frequentist approach is more conservative than the Bayesian one.
This is because in general, for reasonably good ML values
$\tilde{\chi}_{\rm{ML}}^2 \lsim \OO(F)$ and $F > 2$,
\begin{equation}
\int_{\tilde{\chi}_F^2}^{\infty} P^{(F)}(x) dx
 = \int_{\tilde{\chi}_2^2}^{\infty} P^{(2)}(x) dx~
\end{equation}
only for $\tilde{\chi}_F^2 > \tilde{\chi}_2^2$. When looking at
likelihood contours one should thus keep in mind that a point more
than say $3 \si$ away from the ML point is not necessarily ruled
out by data, as we shall show below. In order to establish this,
one has to look at confidence contours, \IE, ask the frequentist's
question. In the following, the term ``likelihood contours'' will
refer to contours drawn in the Bayesian approach, while the term
``confidence contours'' will be reserved for contours coming form
the frequentist point of view.

\section{Results}

\subsection{Adiabatic perturbations}

We first fit CMB data only ($N = 42$) by maximizing $M = 7$
parameters, \IE, the BOOMERanG and Archeops calibration errors,
BOOMERanG beam size error, $n_\SCAL$, $h$, $\OLa$ and the overall
amplitude of the adiabatic spectrum, and we find (Bayesian likelihood
intervals on $\OLa$ alone):
\begin{equation}
\OLa = 0.80 \lims{-0.35}{+0.10} \mbox{ at $2 \si$ $\quad$ and
$\quad$}
            \lims{-0.80}{+0.12}
 \mbox{ at $3 \si$}.
\end{equation}
The asymmetry in the intervals arises because the value of $\OLa$
for our ML model is relatively large. One could achieve a better
precision in determining the ML value of $\OLa$ by using a finer
grid and varying $\omega_\BAR$ as well, which has extensively been
done in the literature and is not the scope of this work.
Moreover, the position of the acoustic peaks in CMB anisotropies
is mainly sensitive to the age of the universe at recombination,
which depends only on $\Om_\MAT h^2$, and to the angular diameter
distance, which depends on $\Om_\MAT$, $\OLa$ and the curvature of
the universe. When the universe is flat, the angular diameter
distance is weakly dependent on the relative amounts of $\Om_\MAT$
and $\OLa$ as soon as $\OLa$ is not too large (see
\EG~Ref.~\cite{melch}). Hence, one can achieve a sufficiently low
value of $\Om_\MAT h^2$ either via a large cosmological constant
or via a very low Hubble parameter, $h \lsim 0.45$.

\begin{figure}[ht]
\centerline{\psfig{file=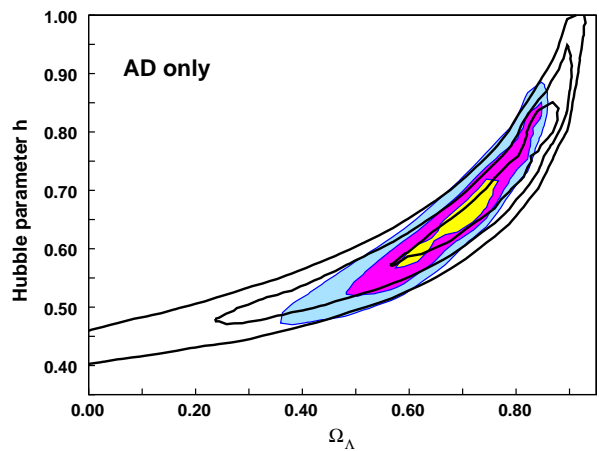,width=9cm}} \caption{Joint
likelihood contours (Bayesian), with CMB only (solid lines,
showing $1 \si$, $2 \si$, $3 \si$ contours) and CMB+LSS (filled)
for purely adiabatic initial conditions.}
\label{f:CL_AD}
\end{figure}

We now include the matter power spectrum $P_\MAT$, assuming
$P_\MAT = b^2 P_\GAL$, where $P_\GAL$ is the observed galaxy power
spectrum and $b$ some unknown bias factor (assumed to be scale
independent), over which we maximize. Inclusion of this data in
the analysis breaks the $\OLa$, $h$ degeneracy, since $P_\MAT$ is
mainly sensitive to the shape parameter $\Gamma \equiv \Om_\MAT
h$. We therefore obtain significantly tighter overall likelihood
intervals for $\OLa$:
\begin{equation}
\OLa = 0.70 \lims{-0.17}{+0.13} \mbox{ at $2 \si$ $\quad$ and
$\quad$}
            \lims{-0.27}{+0.15}
\mbox{ at $3 \si$}.
\end{equation}
We plot joint likelihood contours for $\OLa$, $h$ with AD initial
conditions in \FG \ref{f:CL_AD}.
From the Bayesian analysis, one concludes that CMB and LSS
together require a non-zero cosmological constant at very high
significance, more than $7 \si$ for the points in our grid! Note
that the ML point has a reduced chi-square $\hat{\chi}^2_{F = 56}
= 0.59$, significantly less than $1$.

\begin{figure}[ht]
\centerline{\psfig{file=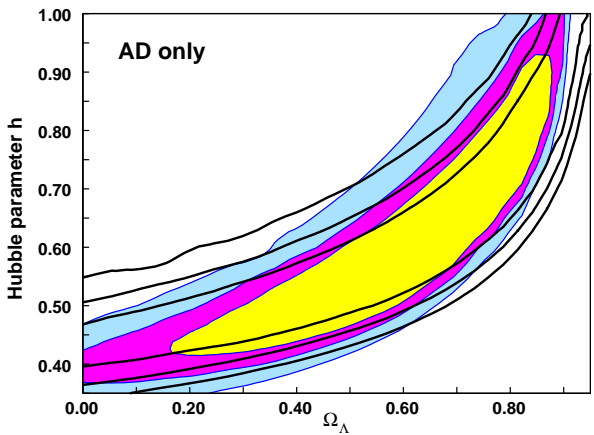,width=9cm}} \caption{Confidence
contours (frequentist) with CMB only (solid lines, $1 \si$, $2
\si$, $3 \si$ contours and $F_\EFF = 31$) and CMB+LSS (filled,
$F_\EFF = 50$) for purely adiabatic initial conditions.}
\label{f:FR_AD}
\end{figure}

\begin{figure}[ht]
\centerline{\psfig{file=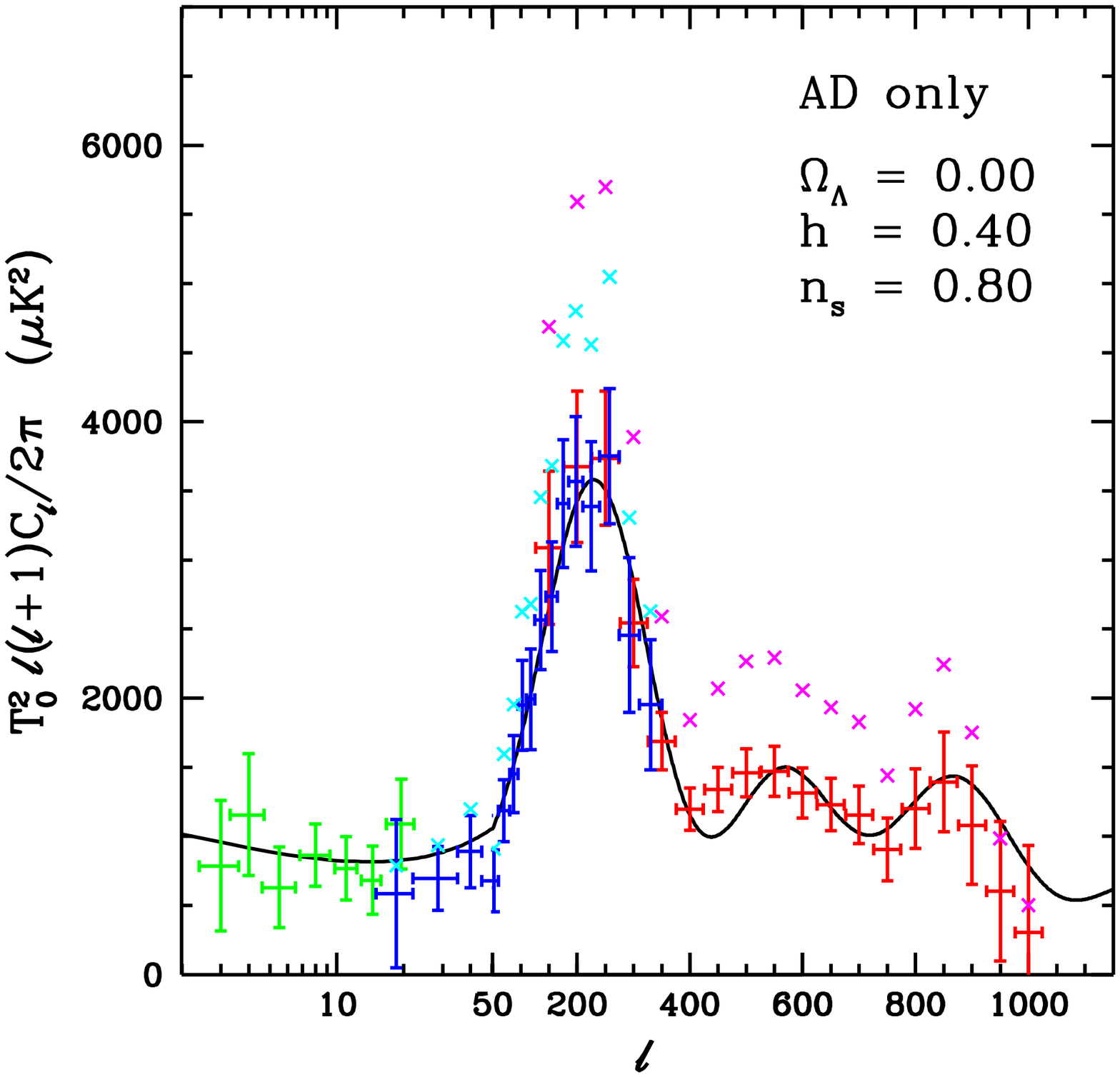,width=6.5cm}}
\centerline{\psfig{file=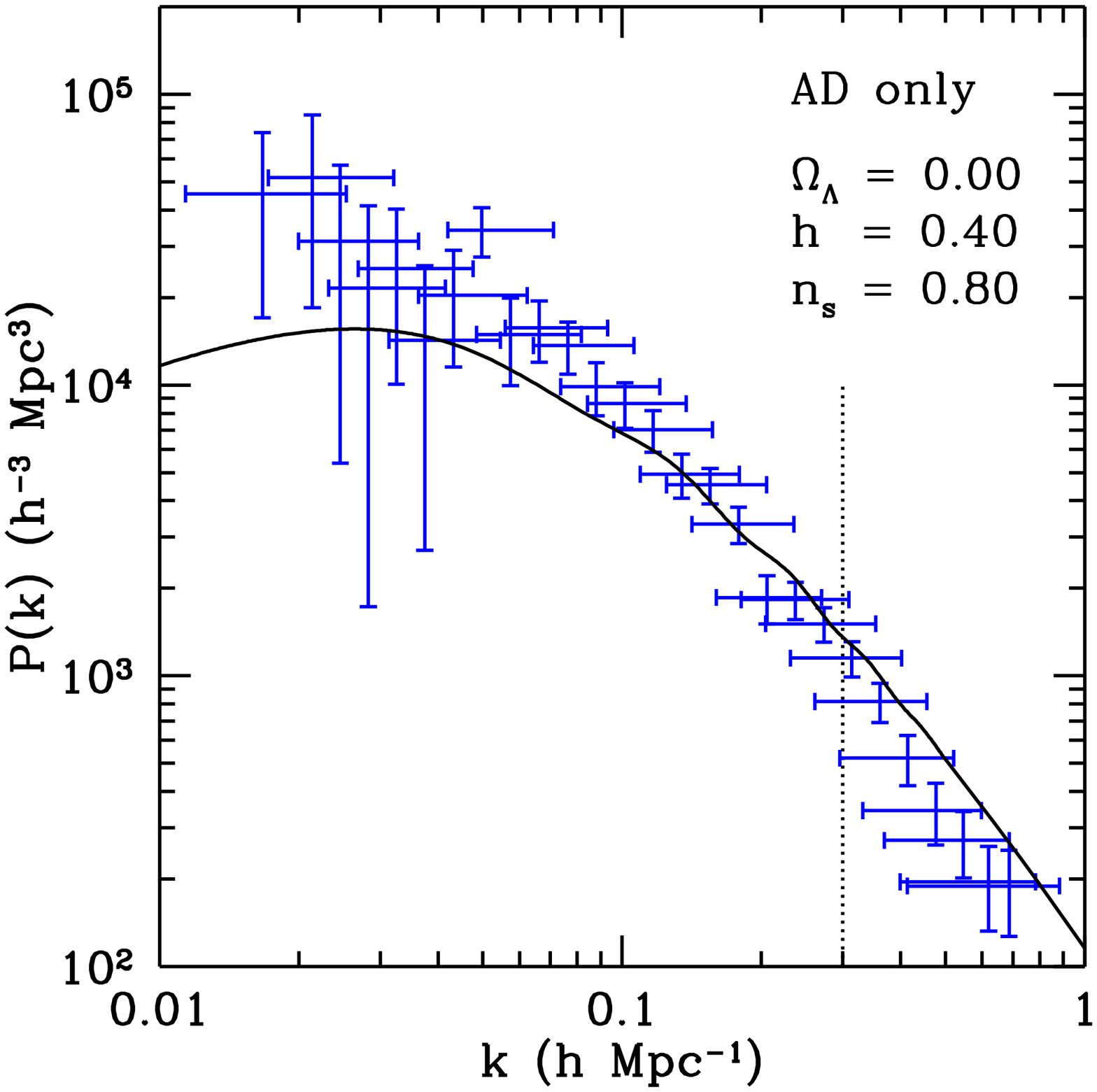,width=6.5cm}} \caption{Best
fit with $\OLa = 0$ and purely AD initial conditions, compatible
with CMB and LSS data within $2 \si$ confidence level. In the
lower panel, only the 2dF data points left of the vertical, dotted
line
--- \IE, in the linear region --- have been included in the analysis. Note the low
first acoustic peak due to the joint effect of the red spectral
index and of the absence of early ISW effect. In this fit, the
calibration of BOOMERanG (red errorbars) and Archeops (blue
errorbars) has been reduced by $34\%$ and $26\%$, respectively.
This is more than 3 times the quoted $1 \si$ calibration errors
for both experiments. To appreciate the difference, we plot the
non recalibrated value of the BOOMERanG and Archeops data points
as light blue and magenta crosses, respectively. In the upper
panel, green errorbars are the COBE measurements. Even though the
fit is ``by eye'' very good, it seems highly unlikely that the
calibration error is so large.}
\label{f:AD_OL0}
\end{figure}

The frequentist analysis, however, excludes a much smaller region
of parameter space (\FG \ref{f:FR_AD}). The frequentist contours
must be drawn for the effective number of dof, \IE, using the
number of effectively independent data points. We can therefore
roughly take into account a $10\%$ correlation, which is the
maximum correlation between data points given in \cite{Arch, Net},
by replacing $F$ by the effective number of dof, $F_\EFF = 0.9 N -
M$, and rounding to the next larger integer (to be conservative).
One could argue that the BOOMERanG and Archeops data points are
not completely independent, since BOOMERanG observed a portion of
the same sky patch as measured by Archeops. This possible correlation
is difficult to quantify, but should not be too important since
the sky portion observed by Archeops is a factor of 10 larger than
BOOMERanG's and therefore we ignore it here. \FG \ref{f:FR_AD} is
drawn with $F_\EFF = 31$ for CMB alone and $F_\EFF = 50$ for
CMB+LSS, but we have checked that our results do not change much
if we use a $5\%$ correlation. It is interesting to note that
there are regions in Fig.~\ref{f:FR_AD} which are excluded with a
certain confidence by CMB data alone but are no longer excluded at
the same confidence when we include LSS data. In other words, it
would seem that taking into account more data and therefore more
knowledge about the universe, does not systematically exclude
more models, \IE, the CMB+LSS contours are not always contained in
the CMB alone contours. This apparent contradiction vanishes when
one realizes that the confidence limits on, \EG, $\OLa$ alone in
the frequentist approach are just the projection of the confidence
contours of \FG \ref{f:FR_AD} on the $\OLa$ axis. One can readily
verify in \FG \ref{f:FR_AD} that the confidence limits for the
combined data-set are always smaller than the ones for CMB data
alone. There are points with $\OLa = 0$ and $h \simeq 0.40$ which
are still compatible within $2 \si$ with both LSS and CMB data, at
the price of pushing somewhat the other parameters. In the best
fit with $\OLa = 0$ shown in Fig.~\ref{f:AD_OL0}, one has to live
with a red spectral index $n_\SCAL = 0.80$. Furthermore, the
calibration of the BOOMERanG and Archeops data points is reduced
in this fit by $34\%$ and $26\%$, respectively, \IE, more than 3
times the quoted $1 \si$ systematic error.

In both cases, it is clear that one can exploit the $\OLa$, $h$
degeneracy to fit CMB data alone with a model having $\OLa = 0$.
For a flat universe like the one we are considering, one has then
to go to a much smaller value of the Hubble parameter than the one
indicated by other measurements, most notably the HST Key
Project~\cite{HST}, which gives $h=0.72 \pm 0.08$. The LSS data
are mainly sensitive to the shape parameter $\Gamma \sim 0.2$.
Hence LSS with $\Om_\MAT = 1.0$ would require an even lower value
of $h$ which is not compatible with CMB. Therefore inclusion of
LSS data tends to exclude any flat model without a cosmological
constant. Summing up, for purely adiabatic initial conditions the
Bayesian approach gives very strong support to $\OLa \neq 0$; in
the more conservative frequentist point of view, while $\OLa \neq
0$ cannot be excluded with very high confidence, present LSS and
CMB data start to be incompatible with a flat universe with
vanishing cosmological constant. These conclusions are in
qualitative agreement with previous
works~\cite{netti,pryke,max,novos,RM,MCMC,Arch2}.

In the next section we investigate the stability of those well
known results with respect to inclusion of non-adiabatic initial
conditions.

\subsection{Isocurvature modes} \label{Icsec2}

\begin{figure}[ht]
\centerline{\psfig{file=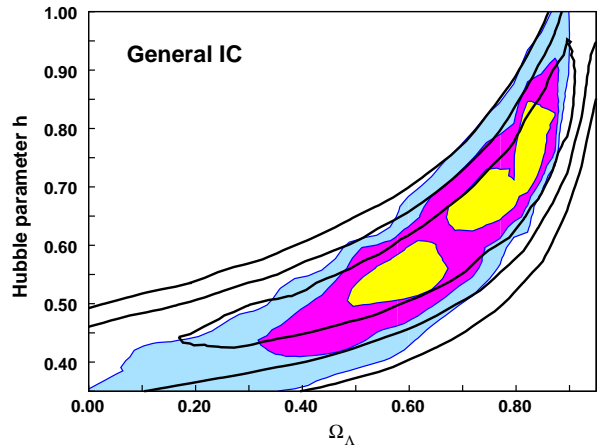,width=9cm}} \caption{Joint
likelihood contours (Bayesian) with general isocurvature initial
conditions, with CMB only (solid lines, $1 \si$, $2 \si$, $3 \si$)
and CMB+LSS (filled). The disconnected $1\si$ region is an
artificial feature due to the grid resolution.}
\label{f:CL_AM}
\end{figure}

\begin{figure}[ht]
\centerline{\psfig{file=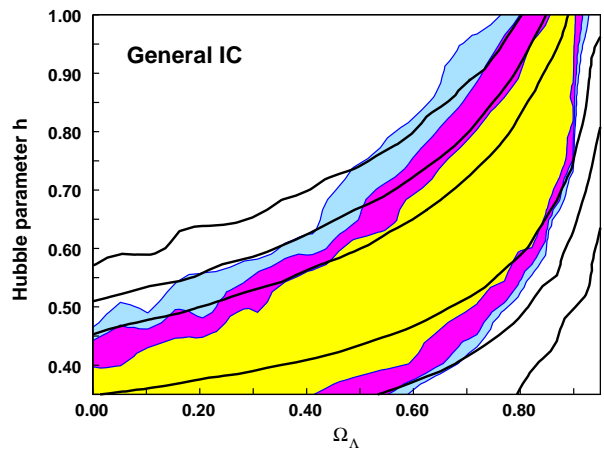,width=9cm}} \caption{Confidence
contours (frequentist) with general isocurvature initial
conditions with CMB only (solid lines, $1 \si$, $2 \si$, $3 \si$
contours $F_\EFF = 22$) and CMB+LSS (filled, $F_\EFF = 41$).}
\label{f:FR_AM}
\end{figure}

\begin{figure}[ht]
\centerline{\psfig{file=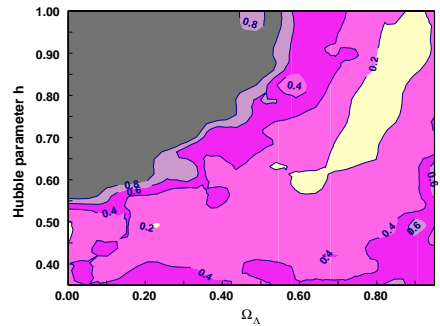,width=6.5cm}}
\caption{Isocurvature content $0.0 \leq \beta \leq 1.0$ of best fit
models with CMB and LSS data. The contours are for $\beta = 0.20,
0.40, 0.60, 0.80$ from the center to the outside.}
\label{f:BETA}
\end{figure}

We now enlarge the space of models by including all possible
isocurvature modes with arbitrary correlations among themselves
and the adiabatic mode as described in the previous section. We
first consider CMB data only and maximize over initial conditions.
The number of parameters increases by nine and the number of dof
decreases correspondingly with respect to the purely AD case
considered above. Likelihood (Bayesian, see Fig.~\ref{f:CL_AM})
and confidence (frequentist, see Fig.~\ref{f:FR_AM}) contours
widen up somewhat along the degeneracy line. The enlargement is
less dramatic than for other parameter choices, see, \EG,
Ref.~\cite{TRD1} where the degeneracy in $\omega_\BAR, h$ was
analyzed. This is partially due to our prior of flatness which
reduces the space of models to the ones which are almost
degenerate in the angular diameter distance. Most of our models
have the first acoustic peak of the adiabatic mode already in the
region preferred by experiments. Hence in most of the fits
isocurvature modes play a modest role, especially in the
parameter regions with large $\OLa$, $h$ (\CF~\FG \ref{f:BETA} and
the discussion below). Nevertheless, because of the $\OLa$, $h$
degeneracy, even a modest widening of the contours along the
degeneracy line results in an important worsening of the
likelihood limits. The ML point does not depart very much from the
purely adiabatic case, but now we cannot constrain $\OLa$ at more
than $1\sigma$ (Bayesian, CMB only):

\begin{equation}
\OLa = 0.85 \lims{-0.35}{+0.05} \mbox{ at $1 \si$, }
\label{e:LL}
\end{equation}
and no limits for $0.0 \leq \OLa \leq 0.95$ at higher confidence.

\begin{figure}[ht]
\centerline{\psfig{file=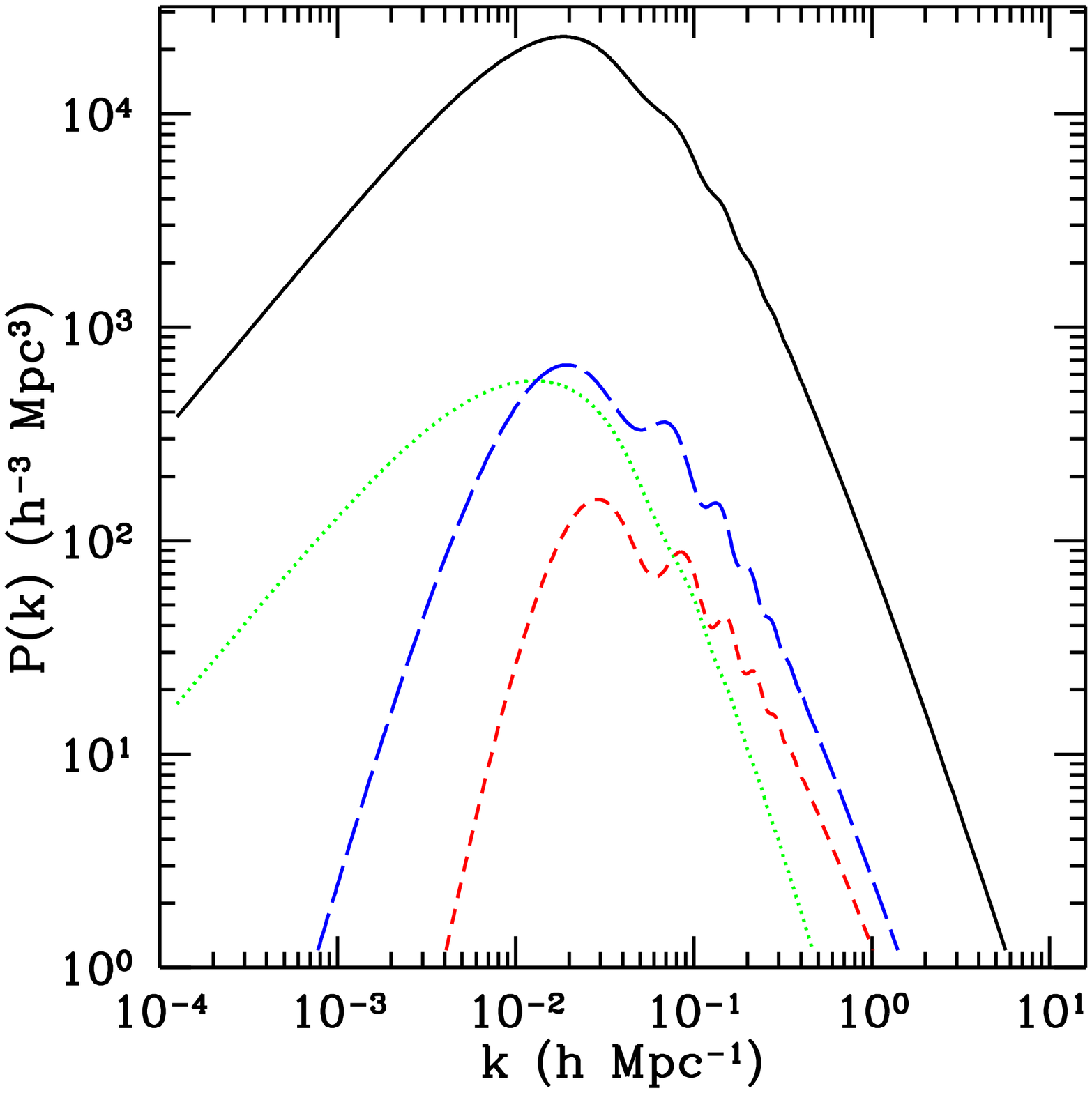,width=6.5cm}}
\centerline{\psfig{file=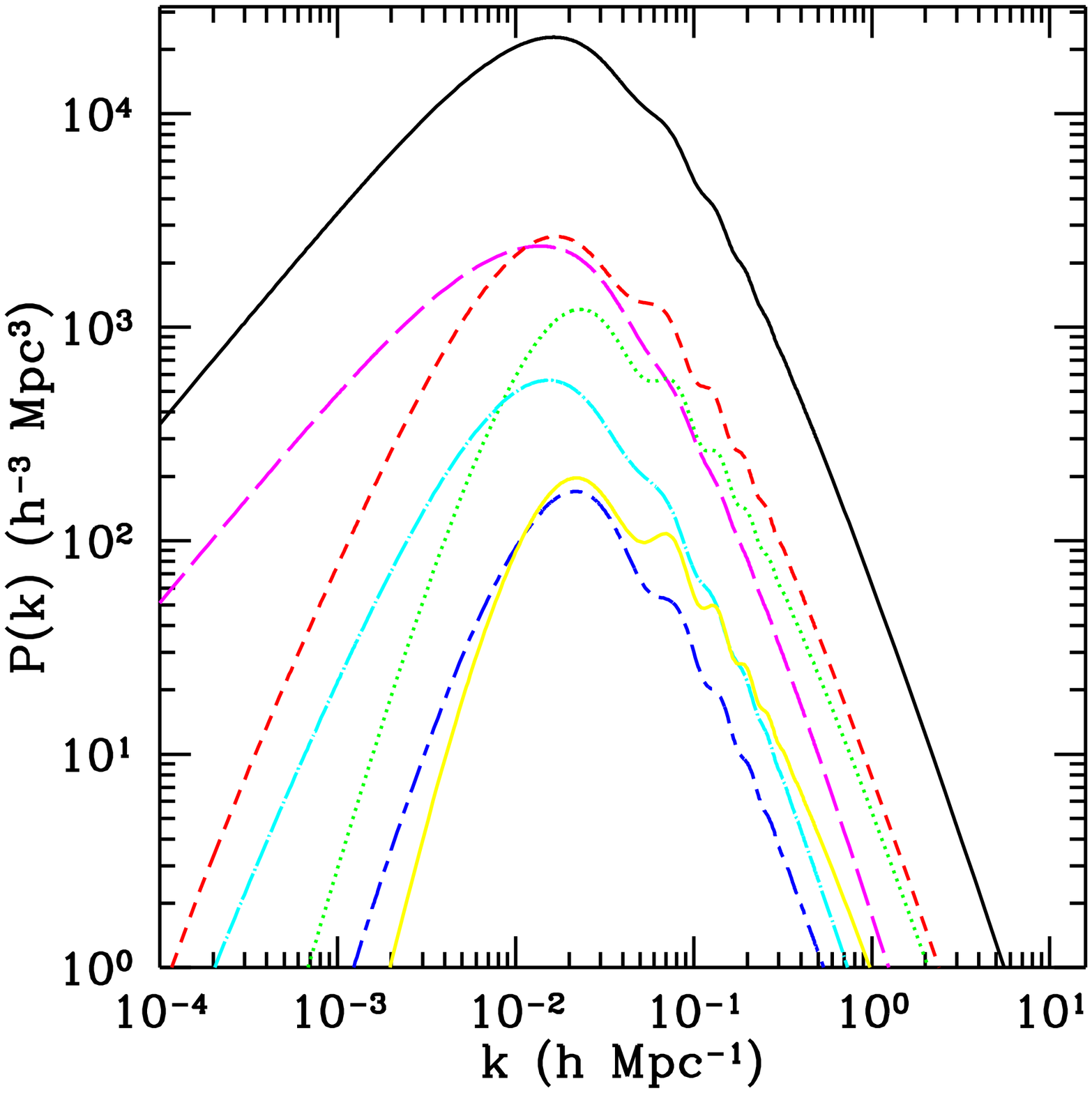,width=6.5cm}}
\caption{Dark matter power spectra of the different auto- (upper
panel) and cross-correlators (lower panel) for a concordance model
with $\OLa = 0.70$, $h = 0.65$, $n_\SCAL = 1.0$, $\omega_\BAR =
0.020$, with the corresponding CMB power spectrum COBE-normalized
(see the text for details). The color codes are as follows: in the
upper panel, AD: solid/black line, CI: dotted/green line, NID:
short-dashed/red line, NIV: long-dashed/blue line; in the lower
panel, AD: solid/black line (for comparison), $<\AD,\CI>$:
long-dashed/magenta line, $<\AD,\NID>$: dotted/green line,
$<\AD,\NIV>$: short-dashed/red line, $<\CI,\NID>$: dot-short
dashed/blue line, $<\CI,\NIV>$: dot-long dashed/light-blue line,
and $<\NID,\NIV>$: solid/yellow line. The adiabatic mode is by far
dominant over all others.}
\label{f:PS_COBE_NORM}
\end{figure}

In Fig.~\ref{f:PS_COBE_NORM} we plot the dark matter power spectra
of the different auto- (upper panel) and cross-correlators (lower
panel) for a concordance model. The norm of each pure mode (AD,
CI, NID, NIV) is chosen such that the corresponding CMB power
spectrum is COBE-normalized. The cross-correlators are normalized
according to totally correlated spectra, \IE
\begin{equation}
A_{(\rm{X},\rm{Y})} = \sqrt{A_{\rm{X}} A_{\rm{Y}}/2}~,
 \label{e:cross_norm}
\end{equation}
where $A_{(\rm{X},\rm{Y})}$ denotes the norm of the
cross-correlator between the modes $X$,$Y$ and $A_{\rm X}$ the
norm of the pure mode $X$. The CMB power spectrum for this set of
cosmological parameters can be found in Ref.~\cite{BMT2}. A
crucial result is that the COBE-normalized amplitude of the AD
matter power spectrum is nearly two orders of magnitude larger
than the isocurvature contribution. The main reason for this is
the amplitude of the Sachs Wolfe plateau which is about
$\frac{1}{3} \Psi$ for adiabatic perturbations and $2 \Psi$ for
isocurvature perturbations. Here $\Psi$ is the gravitational
potential. This difference of a factor of about $36$ in the power
spectrum on large scales is clearly visible in the comparison of
$P_\AD$ and $P_\CI$ (the difference increases at smaller scales).
The case of the neutrino modes is even worse since they start up
with vanishing dark matter perturbations. That the CDM
isocurvature matter power spectrum is much lower than the
adiabatic one has been known for some time (see
\EG~Ref.~\cite{banday}). However, it was not recognized before
that the same holds true for the neutrino isocurvature matter
power spectra as well, and -- more importantly -- that this leads
to a way to break the strong degeneracy among initial conditions
which is present in the CMB power spectrum alone.

\begin{figure}[ht]
\centerline{\psfig{file=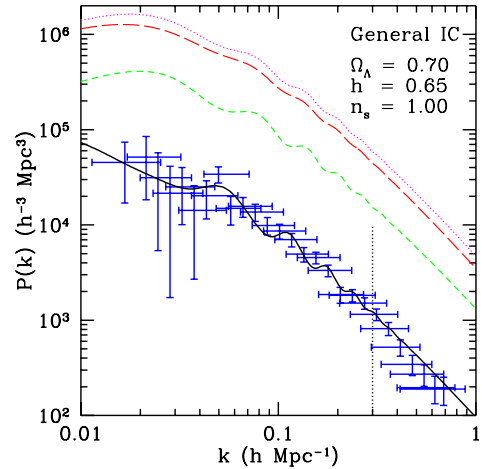,width=6.5cm}}
\caption{Concordance model fit with general IC and LSS data only.
The total spectrum (solid/black) is the result of a large
cancellation of the purely AD part (long-dashed/red) by the large,
negative sum of the various correlators (dotted/magenta, plotted
in absolute value). The short-dashed/green curve is the sum of the
three pure isocurvature modes, CI, NID and NIV. Note that the
resulting total spectrum is less than one tenth of the purely
adiabatic part.}
\label{f:AM_PS_ONLY}
\end{figure}

In an analysis with general initial conditions including LSS data
only we obtain very broad likelihood and confidence contours which
exclude only the lower right corner of the $(\OLa, h)$ plane. In
contrast to the CMB power spectrum, the matter power spectrum can
be fitted with extremely high adiabatic and isocurvature
contributions, which are then typically cancelled by large
anti-correlations between the spectra. This behavior is
exemplified for a model with general IC and $\OLa = 0.70$, $h =
0.65$, $n_\SCAL = 1.0$ in Fig.~\ref{f:AM_PS_ONLY}. The best fits
with LSS data only are dominated by large isocurvature
cross-correlations. Clearly, the resulting CMB power spectrum is
highly inconsistent with the COBE data. Hence such ``bizarre''
possibilities are immediately ruled out once we include CMB
data. Conversely, moderate isocurvature contributions can help
fitting the CMB data, and do not influence the matter power
spectrum, which is completely dominated by the AD mode alone.
Combining CMB and LSS data (see \FG \ref{f:CL_AM}) we find now
(Bayesian, mixed IC):
\begin{equation}
\OLa = 0.65 \lims{-0.25}{+0.22} \mbox{ at $2 \si$ $\quad$ and
$\quad$}
            \lims{-0.48}{+0.25}
\mbox{ at 3$\si$}.
\end{equation}
The likelihood limits are larger than for the purely adiabatic
case but it is interesting that the Bayesian analysis still
excludes $\OLa = 0$ at more than $3 \si$ even with general initial
conditions, for the class of models considered here. Because of
the above explained reason, the widening of the limits is not as
drastic as one might fear. Therefore, combination of CMB and LSS
measurements turn out to be an ideal tool to constrain the
isocurvature contribution to the initial conditions.

\begin{figure}[ht]
\centerline{\psfig{file=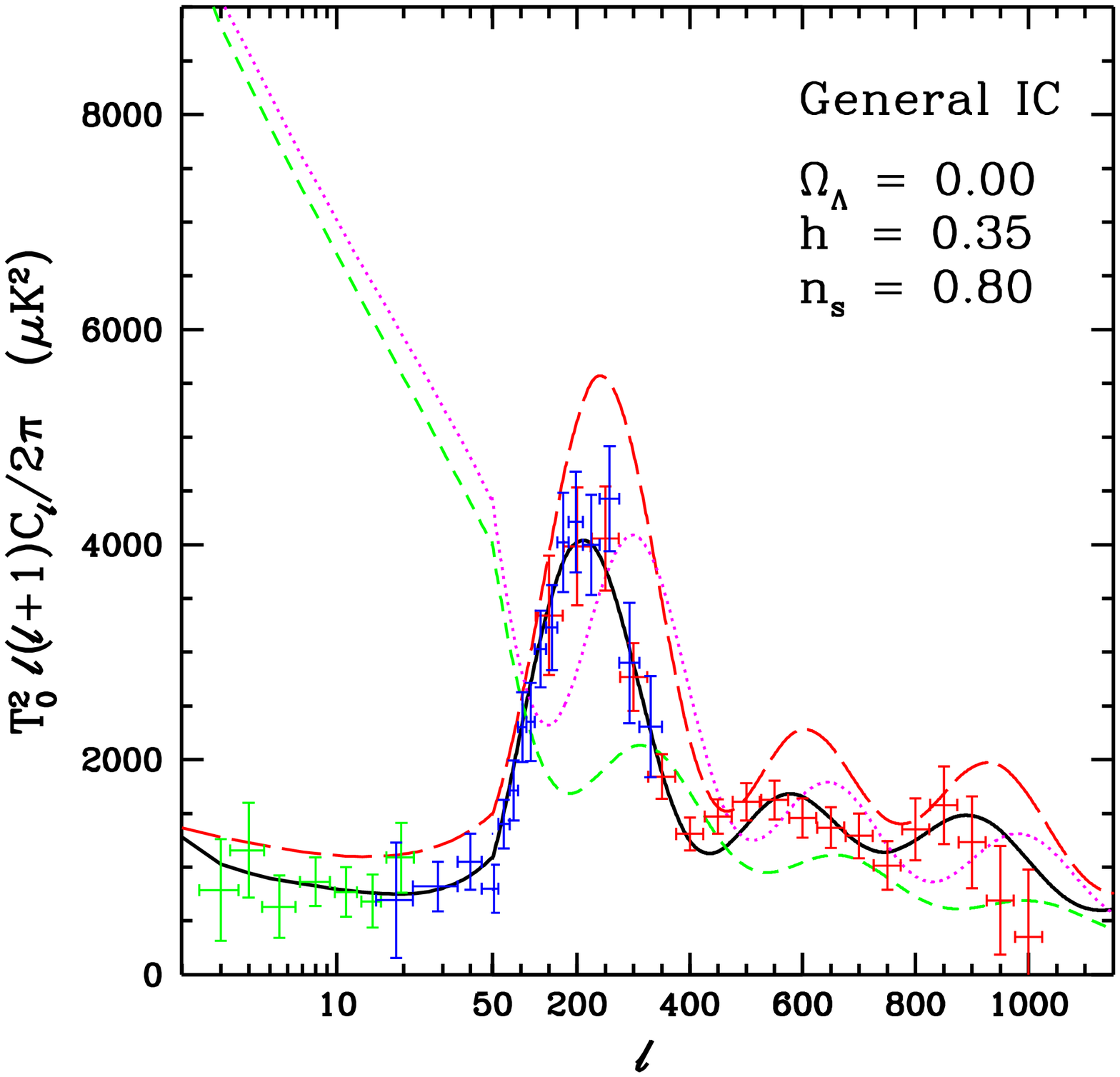,width=6.5cm}}
\centerline{\psfig{file=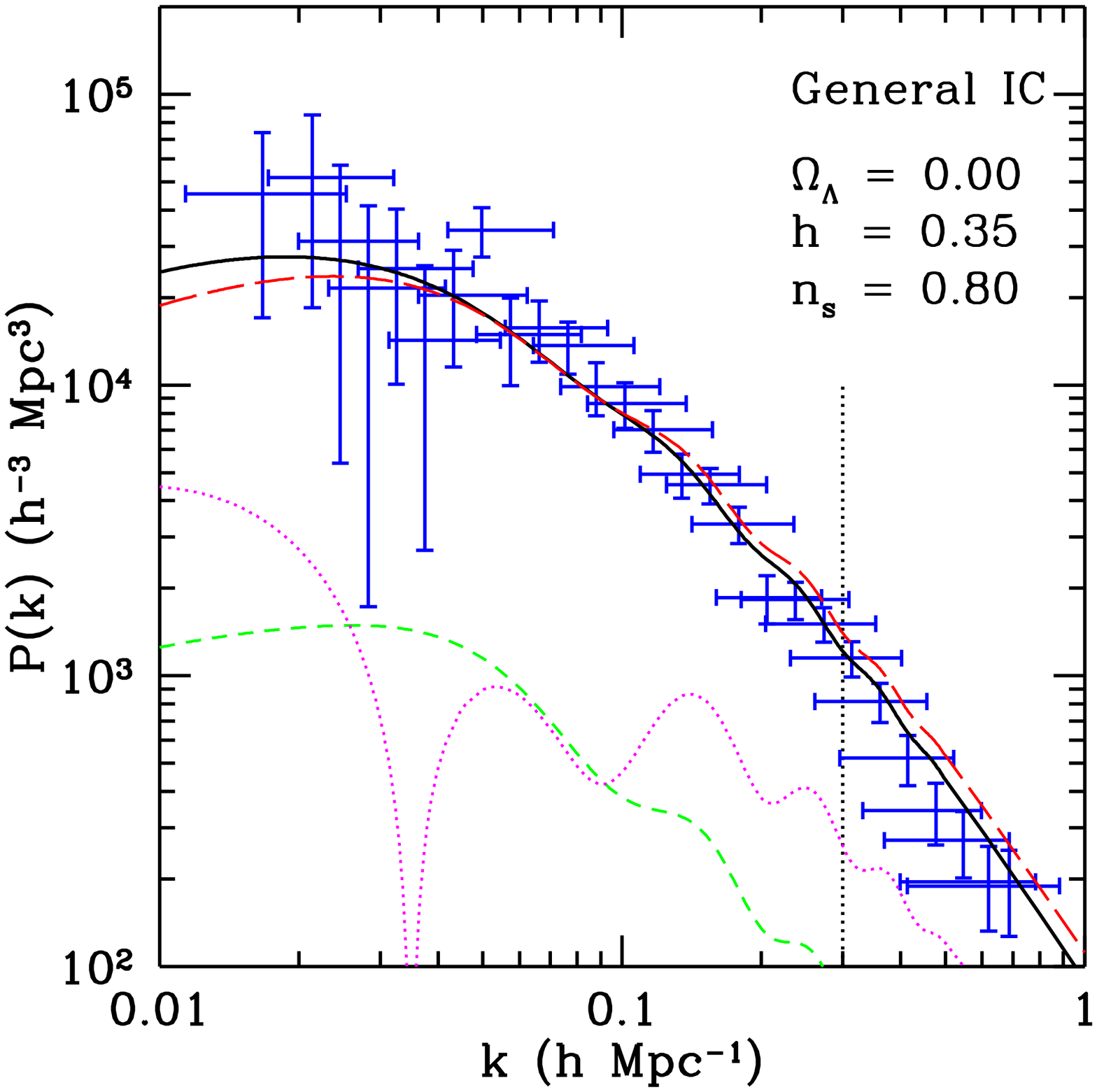,width=6.5cm}} \caption{Best
fit with general IC and $\OLa = 0$. As for purely AD, even with
general IC the absence of the cosmological constant suppresses in
an important way the height of the first peak. In both panels we
plot the best total spectrum (solid/black), the purely AD
contribution (long-dashed/red), the sum of the pure isocurvature
modes (short-dashed/green) and the sum of the correlators
(dotted/magenta, multiplied by $-1$ in the upper panel and in
absolute value in the lower panel). The matter power spectrum is
completely dominated by the AD mode, while the correlators play an
important role in cancelling unwanted contributions in the CMB
power spectrum at the level of the first peak and especially in
the COBE region. For this model we have $\beta = 0.39$, while the
BOOMERanG and Archeops calibrations are reduced by $28\%$ and
$12\%$, respectively.}
\label{f:AM_OL0}
\end{figure}

From the frequentist point of view, one notices that the region in
the $\OLa, h$ plane which is incompatible with data at more than
$3 \si$ is nearly independent on the choice of initial conditions
(compare \FG \ref{f:FR_AD} and \FG \ref{f:FR_AM}). Enlarging the
space of initial conditions seemingly does not have a relevant
benefit on fitting present-day data with or without a cosmological
constant. The reason for this is that the (COBE-normalized) matter
power spectrum is dominated by its adiabatic component and
therefore the requirement $\Om_\MAT h \sim 0.2$ remains valid.  In
\FG \ref{f:AM_OL0} we plot the best fit model with general initial
conditions and $\OLa = 0$. We summarize our likelihood and
confidence intervals on $\OLa$ (this parameter only) in Table 1.

In Fig.~\ref{f:BETA} we plot the isocurvature contribution to the
best fit models with CMB and LSS in terms of the parameter $\beta$
defined in Eq.~(\ref{e:beta}). The best fit with $\OLa = 0$ has an
isocurvature contribution of about $40\%$.

It would be of great interest to investigate whether the result
$\OLa \neq 0$ is robust with respect to addition of general
initial conditions in an open universe~\cite{TRDprep}.

\begin{table*}
\caption{Results for the likelihood (Bayesian) and confidence
(frequentist) intervals for $\OLa$ alone (all other parameters
maximized). A bar, $-$, indicates that at the given
likelihood/confidence level the analysis cannot constraint $\OLa$
in the range $0.0 \leq \OLa \leq 0.95$. Where the quoted interval
is smaller than our grid resolution, an interpolation between
models has been used.}
\begin{tabular}{|c|c |c c c | c c c | c c|}
\hline \hline
 \multicolumn{10}{|c|}{AD only}  \\
 \hline
 \multicolumn{2}{|c}{ }        & \multicolumn{3}{|c}{Bayesian \footnotemark[3]} \rule{0pt}{4ex}
                              & \multicolumn{3}{|c}{Frequentist \footnotemark[4]} \rule{0pt}{4ex} &  &\\
 Data-set  & $\quad \OLa$ \quad & $1\si$ & $2\si$  & $3\si$ & $1\si$ & $2\si$
          & $3\si$ & $F$ & $\chi^2/F$\\
\hline CMB\footnotemark[1]+flatness \rule{0pt}{4ex}
          & $0.80$& $\lims{-0.08}{+0.08}$
                  & $\lims{-0.35}{+0.10}$
                  & $\lims{\quad -}{+0.12}$
                  & $ < 0.93 $
                  & $ - $
                  & $ - $
                  & $35 $    & $0.58$\\
CMB\footnotemark[1] +LSS\footnotemark[2]+flatness \rule{0pt}{4ex}
          & $0.70$& $\lims{-0.05}{+0.05}$
                  & $\lims{-0.17}{+0.13}$
                  & $\lims{-0.27}{+0.15}$
                  & $ 0.15 < \OLa < 0.90 $
                  & $ \quad < 0.92 $
                  & $\quad < 0.92 $
                  & $56 $    & $0.59$\\
\hline
\multicolumn{10}{|c|}{General IC}  \\
\hline CMB\footnotemark[1]+flatness \rule{0pt}{4ex}
          & $0.85$& $\lims{-0.35}{+0.05}$
                  & $ - $
                  & $ - $
                  & $ - $
                  & $ - $
                  & $ - $
                  &  $26 $   & $0.74$\\
CMB\footnotemark[1] +LSS\footnotemark[2]+flatness \rule{0pt}{4ex}
          & $0.65$& $\lims{-0.10}{+0.15}$
                  & $\lims{-0.25}{+0.22}$
                  & $\lims{-0.48}{+0.25}$
                  & $ < 0.90 $
                  & $ < 0.92 $
                  & $\quad < 0.95 $
                  &  $47 $   & $0.67$\\
\hline \hline

\end{tabular}
\footnotetext[1]{~COBE, BOOMERanG and Archeops data.}
\footnotetext[2]{~2dF data.}
\footnotetext[3]{~Likelihood interval.}
\footnotetext[4]{~Region not excluded by data with given confidence.}
\end{table*}

\section{Conclusions}

The conclusions of this work are threefold. The first one is not
new, but seems to be dangerously forgotten in recent cosmological
parameters estimation literature: namely that likelihood contours
cannot be used as ``exclusion plots''. The latter are usually
substantially wider, less stringent. A more rigorous possibility
are frequentist probabilities, which however suffer from the
dependence on the number of really independent measurements which
is often very difficult to come by.

Secondly, we have found that in COBE-normalized fluctuations, the
matter power spectrum has negligible isocurvature contributions
and is essentially given by the adiabatic mode. Hence the
shape of the observed matter power spectrum still requires
$\Om_\MAT h \simeq 0.2$, independent of the choice of initial
conditions. Due to this behavior, the condition $\Om = \OLa +
\Om_\MAT = 1$ requires either a cosmological constant or a very
small value for the Hubble parameter, independently from the
isocurvature contribution to the initial conditions.

The third conclusion from our work are the following results for
the presence of a cosmological constant: For flat models, a
likelihood (Bayesian) analysis strongly favors a non-vanishing
cosmological constant. Even if we allow for isocurvature
contributions with arbitrary correlations, a vanishing
cosmological constant is still excluded at more than $3 \si$. If
we would allow for open models, a significant contribution from
the NIV mode which has the first acoustic peak at $\ell = 170$ in
flat models, possibly could at the same time give a good fit to
CMB data and allow for the observed shape parameter $\Ga$ with a
reasonable value of $h$. For technical reasons we shall study this
case in a forthcoming paper~\cite{TRDprep}.

The situation changes considerably in the frequentist approach.
There, even for purely adiabatic models, $\OLa = 0$ is still within
$3 \si$ for a value of $h \le 0.48$ which is marginally
defendable.  The conclusion does not change very much when we
allow for generic initial conditions.

\acknowledgments

We thank Alessandro Melchiorri, who participated in the beginning
of this project, and Alain Blanchard for stimulating discussions.
RT was partially supported by the Schmidheiny Foundation. This
work is supported by the Swiss National Science Foundation and by
the European Network CMBNET.

 \end{document}